\documentclass[aps,preprint,epsfig,rotate]{revtex4}
\usepackage{graphicx}
\usepackage{bm}
\usepackage{epsfig}

\begin{document}

\title{One-photon annihilation of the bound electron-positron pair in 
       the field of atomic nuclei} 

\author{Alexei M. Frolov}
 \email[E--mail address: ]{alex1975frol@gmail.com}  

\affiliation{Department of Applied Mathematics \\
 University of Western Ontario, London, Ontario N6H 5B7, Canada}

\date{\today}

\begin{abstract}

One-photon annihilation of the bound electron-positron pair in the field of 
atomic nuclei is investigated. In our analysis the central atomic nucleus is 
assumed to be an infinitely heavy, positively charged and point `particle' 
with the electric charge $Q e$. The explicit formulas for the both 
annihilation cross section and one-photon annihilation rate $\Gamma^{(b)}_{1 
\gamma}$ are derived and discussed. We also derive and discuss the closed 
formula which describes the angular distribution of the emitted photons. All 
our formulas can directly be used to evaluate probabilities of the actual 
one-photon annihilation in various atoms, ions and light molecular ions which 
also contain one (or two) positron(s). 

\end{abstract}

\maketitle

\newpage
\section{Introduction}

In this study we investigate one-photon annihilation of the bound electron-positron pair 
(or $(e^{-}, e^{+})-$pair, for short) in the Coulomb field of a central, infinitely heavy, 
positively charged atomic nucleus. Our main focus will be on non-relativistic annihilation 
from bound states of light atomic systems. In respect to this, we shall not consider 
one-photon annihilation `in-flight' in heavy atoms with $Q \ge 80$. Annihilation of the 
bound $(e^{-}, e^{+})-$pair in various atomic systems is of great interest in the Stellar 
and Solar astrophysics (see, e.g., \cite{Dra1} - \cite{Panth2} and references therein). 
Indeed, this process often proceeds in hydrogen and helium atoms from Solar and Stellar 
photoshperes. Therefore, it is important to determine its probability (or rate) correctly 
and accurately. To achieve this goal we need to derive the exact formulas for one-photon 
annihilation rate of the bound $(e^{-}, e^{+})-$pair in various atomic systems. 

The problem of one-photon annihilation was discussed in a number of earlier studies (see, 
e.g., \cite{Heitl} and \cite{OurHPs}). However, Heitler in his book \cite{Heitl} reduced 
the original problem of one-photon annihilation of the $(e^{-}, e^{+})-$pair in atoms to 
calculation of the usual `atomic' photodetachment. The only difference with actual 
one-photon annihilation was a replacement of the final electron by an incoming positron. 
Results of such an approximate analysis lead to relatively large deviations in numerical 
evaluations of the total probability of one-photon annihilation for the bound $(e^{-}, 
e^{+})-$pair for light atoms, i.e., for the systems which are of main interest in this 
study. 

In our earlier paper \cite{OurHPs} the same one-photon annihilation rate in the positronium 
hydride HPs was approximately evaluated by using the known analytical formula for one-photon 
annihilation rate $\Gamma_{1 \gamma}$ in the three-body positronium ion Ps$^{-}$. However, 
it was $a$ $priori$ clear that the overall accuracy of such an approximation for the 
four-body HPs system cannot be good, since the three-body Ps$^{-}$ ion does not include 
any heavy `proton' (or central nucleus). For this reason our recently published papers 
\cite{Fro2022}, \cite{Fro2025} do not contain any evaluation of the one-photon annihilation 
rate(s), while the two-, three-, four- and five-photon annihilation rates have been
determined in \cite{Fro2025} quite accurately. This indicates clearly that even a two years 
ago we did not have the correct QED-based formula which can be used in such calculations. 

Note also that the rate of one-photon annihilation of the bound electron-positron pair at 
the heavy nucleus in few- and many-electron atomic system was defined in \cite{OurHPs} as 
the second one-photon annihilation rate $\Gamma^{(b)}_{1 \gamma}$. In this study we want 
to correct inaccuracies from \cite{OurHPs} and derive the correct and exact QED formula 
for one-photon annihilation rate, which corresponds to annihilation of the bound $(e^{-}, 
e^{+})-$pair in the Coulomb field of a central, very heavy atomic nucleus. At this stage 
of our calculations we shall neglect by the nuclear recoil, i.e., we restricted ourselves 
here only to atomic systems with the infinitely heavy nuclei. Then we derive the universal 
formula for the one-photon annihilation rate which can be applied to few-electron atomic 
systems with arbitrary nuclei (or quasi-nuclei) which also contain positrons.  

This paper has the following structure. In the next two Sections we derive the closed 
analytical QED formulas for the squared scattering amplitude and unpolarized cross section, 
which are used later to describe the one-photon annihilation of electron-positron pair in 
the Coulomb field of a central, infinitely heavy atomic nucleus which has positive electric 
charge $Q e (> 0)$. Angular distribution of the emitted photons is investigated in the 
fourth Section. In Section V we discuss the low-energy limit of the cross section of 
one-photon annihilation of the electron-positron pair in the Coulomb field of an infinitely 
heavy atomic nucleus ($Q e$). Applications of our newly derived formulas to actual atomic 
systems can be found in Section VI. Concluding remarks can be found in the last (seventh) 
Section.  

\section{Basic QED calculations} 
    
The two Feynman graphs in momentum space which describe (in the lowest-order) the one-photon 
annihilation of the electron-positron pair in the Coulomb field of a heavy nucleus are shown 
in Fig.1. By applying the Feynman rules of Quantum Electrodynamics (see, e.g., \cite{AB} and 
\cite{Grein}) one easily obtains the explicit expression for the $S-$matrix element in the 
coordinate space. In relativistic units, where $\hbar = 1$ and $c = 1$ the explicit expression 
for the $S_{fi}$ matrix element is written in the following form 
\begin{eqnarray}
 &&S_{fi} = - e^{2} \int \int d^{4}x d^{4}y \; \sqrt{\frac{m^{2}_{0}}{E_{+} E_{-} 
 V^{2}}} \; \sqrt{\frac{4 \pi}{2 \omega V}} \; \Bigl\{ \tilde{v}(p_{+},s_{+}) 
 \Bigl[ (-\imath \epsilon) \; \Bigl(\exp(-\imath k x) + \exp(\imath k x)\Bigr) \; 
 \times \nonumber \\
 && \; \; \imath \; S_{F}(x - y) \; (-\imath \gamma^{0}) \; A^{Coul}_{0}(y) + 
 (-\imath \gamma^{0}) \; A^{Coul}_{0}(x) \; \imath \; S_{F}(x - y) \; 
 (-\imath \epsilon) \; \Bigl(\exp(-\imath k y) + \nonumber \\
 && \; \; \exp(\imath k y)\Bigr) \; u(p_{-},s_{-}) \Bigr\} \; \; , \; \label{Sfi1}
\end{eqnarray} 
where $m_0$ is the electron's mass at rest, $\gamma^{0}$ is the corresponding Dirac 
$\gamma-$matrix, $\imath = \sqrt{-1}$ is the imaginary unit, while the notation 
$A^{Coul}_{0}(x)$ designates the static Coulomb potential with the following components 
\begin{eqnarray}
 A^{Coul}_{0}(x) = A_{0}({\bf x}) = - \frac{Q e}{\mid {\bf x} \mid}  
 = - \frac{Q e}{\mid r \mid} \; \; , \; \; {\rm and} \; \; {\bf A}^{Coul}(x) = 
 {\bf A}({\bf x}) = 0 \; . \; \label{Coul}
\end{eqnarray}
The three-dimensional (or spatial) Fourier transform of this static potential reads
\begin{eqnarray}
  A_{0}({\bf q}) = - Q e \; \int d^{4}x \frac{\exp(-\imath {\bf q} \cdot {\bf 
  x})}{\mid {\bf x} \mid} = \frac{- Q e}{\mid {\bf q} \mid^{2}} \; \; \; \; 
  {\rm and} \; \; \; {\bf A}({\bf q}) = 0 \; . \; \label{Coulom}
\end{eqnarray}
Also, in Eq.(\ref{Sfi1}) the notation $S_{F}(x - y)$ means the following function  
\begin{eqnarray}
  S_{F}(x - y) = \frac{\exp[- \imath p (x - y)]}{p - m_0 + \imath \varepsilon}
  \; \; \; , \; \; \nonumber 
\end{eqnarray} 
where $\varepsilon$ is a very small positive number. 

At the second step of the standard QED procedure we have to perform integration over 
the $x$ and $y$ variables in Eq.(\ref{Sfi1}). This leads to the following expression 
for the $S-$matrix element in the momentum representation 
\begin{eqnarray}
 &&S_{fi} =  Q e^{3} \; [2 \pi \delta(\omega - E_{-} - E_{+})] \; 
 \sqrt{\frac{m^{2}_{0}}{E_{+} E_{-} V^{2}}} \; \; \sqrt{\frac{4 \pi}{2 \omega V}} 
 \; \; \frac{4 \pi}{\mid {\bf q}\mid^{2}} \; \times \nonumber \\
 && \Bigl\{\tilde{v}(p_{+},s_{+}) \Bigl[ (-\imath \epsilon) \; \frac{\imath}{- 
 p_{+} + k - m_0} \;  (-\imath \gamma^{0}) + (-\imath \gamma^{0}) \; 
 \frac{\imath}{p_{-} - k - m_0} \; (-\imath \epsilon) \Bigr] \; 
 u(p_{-},s_{-})\Bigr\} \; \; , \; \label{Sfi2}
\end{eqnarray}
where $p_{+} = (E_{+}, {\bf p}_{+}), p_{-} = (E_{-}, {\bf p}_{-})$ and $k = (\omega, 
{\bf k})$ are the corresponding four-vectors of positron, electron and photon, 
respectively. Also, the electron's and positron's bi-spinors in Eq.(\ref{Sfi1}) are 
designated by the notations $u(p_{-},s_{-})$ and $v(p_{+},s_{+})$, respectively.  

In relativistic units, where $c = 1$ and $\hbar = 1$, we always have 
$p_{\pm} \cdot p_{\pm} = E^{2}_{\pm} - {\bf p}^{2}_{\pm} = m^{2}_{0}$ and $k \cdot 
k = \omega^{2} - {\bf k}^{2} = 0$. From this equation one finds the following 
formula for the $\mid S_{fi} \mid^{2} \frac{d^{3}\vec{\omega}}{(2 \pi)^{3}}$ 
differential, which takes the form 
\begin{eqnarray}
 & &\mid S_{fi} \mid^{2} V \frac{d^{3}\vec{\omega}}{(2 \pi)^{3}} = Q^{2} e^{6} 
 \; [2 \pi \delta(\omega - E_{+} - E_{-})]^{2} \; \frac{32 \; \pi^{3} \; 
 m^{2}_{0} \; V \; \omega^{2} \; d\omega \; d{\bf n}_{f}}{8 \; \pi^{3} \; 
 \omega \; E_{+} \; E_{-} \; V^{3}} \; \; \frac{1}{\mid {\bf q}\mid^{4}} \; 
 \bar{F}(p_{+}, p_{-}; k) \nonumber \\
 &=& Q^{2} e^{6} \; [2 \; \pi \; \delta(0) \; 2 \; \pi \; \delta(\omega - 
 E_{-} - E_{+})] \; \frac{4 \; m^{2}_{0} \; \omega \; d\omega \; 
 d{\bf n}_{f}}{E_{+} \; E_{-} \; V^{2}} \; \; \frac{1}{\mid {\bf q} \mid^{4}} 
 \; \bar{F}(p_{+}, p_{-}; k) \; \; , \; \nonumber \\
 &=& 8 \pi Q^{2} e^{6} \; T \; \delta(\omega - E_{-} - E_{+}) \; 
 \frac{m^{2}_{0} \; \omega \; d\omega \; d{\bf n}_{f}}{E_{+} \; E_{-} 
 \; V^{2}} \; \; \frac{1}{\mid {\bf q} \mid^{4}} \; 
 \bar{F}(p_{+}, p_{-}; k) \; \; , \; \label{Sfi3} 
\end{eqnarray}
where we replace $2 \pi \delta(0) \Rightarrow T$ (see, e.g., \cite{Grein}), ${\bf 
n}_f$ is the unit vector which is directed along the direction of the emitted (or 
out-going) photon ${\bf k}$, while the function $\bar{F}(p_{+}, p_{-}; k)$ is the 
squared scattering amplitude modulus which is explicitly written in the form 
\begin{eqnarray}
 & & \bar{F}(p_{+}, p_{-}; k) = \Bigl| \tilde{v}(p_{+},s_{+}) \Bigl[ \epsilon 
 \; \frac{1}{- p_{+} + k - m_0} \; \gamma^{0} + \gamma^{0} \; \frac{1}{p_{-} 
 - k - m_0} \; \epsilon \Bigr] u(p_{-},s_{-}) \Bigr|^{2} \; \nonumber \\
 &=& \Bigl\{\tilde{v}(p_{+},s_{+}) \Bigl[ \epsilon \; \frac{1}{- p_{+} + k 
 - m_0} \; \gamma^{0} + \gamma^{0} \; \frac{1}{p_{-} - k - m_0} \; 
 \epsilon \Bigr] \; \nonumber \\
 & & u(p_{-},s_{-}) \tilde{u}(p_{-},s_{-}) \Bigl[ \gamma^{0} \; 
 \frac{1}{- p_{+} + k - m_0} \;  \epsilon + \epsilon \; \frac{1}{p_{-} 
 - k - m_0} \; \gamma^{0} \Bigr] v(p_{+},s_{+}) \Bigr\} \; \; , 
 \; \label{Sfi33}
\end{eqnarray} 
where all notations for the electron and positron bi-spinors, Green's functions, 
etc, are standard and coincide with the corresponding notations used in 
\cite{Grein}. 

Now, the formula for the unpolarized cross section is written in the form 
\begin{eqnarray}
 d\sigma = \int \frac{\mid S_{fi} \mid^{2}}{T \frac{v_{+}}{V} \frac{v_{-}}{V}} 
 \frac{d^{3}\vec{\omega}}{(2 \pi)^{3}} = 8 \pi Q^{2} \alpha^{3} \; 
 \Theta(\omega - E_{-} - E_{+}) \; \frac{m^{2}_{0} \; \omega \; d\omega \; 
 d{\bf n}_{f}}{v_{+} \; E_{+} \; v_{-} \; E_{-}} \; \; \frac{1}{\mid {\bf q} 
 \mid^{4}} \; \bar{F}(p_{+}, p_{-}; k) \; \; , \; \label{Sfi35} 
\end{eqnarray}
where $v_{+} = \mid {\bf v}_{+}  - {\bf V}_{N} \mid$ and $v_{-} = \mid {\bf v}_{-} 
- {\bf V}_{N} \mid$ are the positron's and electron's velocities, respectively. 
Note that each of them are taken in respect to the central heavy nucleus. These two 
velocities are directed at the central atomic nucleus and they cannot be replaced 
by the relative electron-positron velocity ${\bf v}_{rel} = \mid {\bf v}_{-} - 
{\bf v}_{+} \mid$. In other words, the one-photon annihilation of electron-positron 
pair in the central Coulomb field of an atomic nucleus is a three-particle process 
which equally includes one positron $e^{+}$, one electron $e^{-}$ and central atomic 
nucleus $Q e$.  

In real applications the unpolarized (differential) cross section, Eq.(\ref{Sfi35}), 
must be averaged over the spin directions $s_{-}$ and $s_{+}$ and summed over the 
photon polarizations $\lambda$. This gives the additional numerical factor 2. Finally, 
one finds from Eq.(\ref{Sfi35}) the following expression for the unpolarized cross 
section    
\begin{eqnarray}
 d\sigma = \int \frac{\mid S_{fi} \mid^{2}}{T \frac{v_{+}}{V} \frac{v_{-}}{V}} 
 \frac{d^{3}\vec{\omega}}{(2 \pi)^{3}} = 16 \pi Q^{2} \alpha^{3} \; \Theta(\omega 
 - E_{-} - E_{+}) \; \frac{m^{2}_{0} \; \omega \; d\omega \; d{\bf n}_{f}}{\mid 
 {\bf p}_{+} \mid \mid {\bf p}_{-} \mid} \; \; \frac{1}{\mid {\bf q} \mid^{4}} 
 \; \tilde{F}(p_{+}, p_{-}; k) \; \; , \; \label{Sfi351} 
\end{eqnarray}
where we also used the fact that in relativistic units: ${\bf v}_{+} \; E_{+} = {\bf p}_{+}$ 
and ${\bf v}_{-} \; E_{-} = {\bf p}_{-}$. Also, it is important to note that the right-hand 
sides of the last two equations (Eqs.(\ref{Sfi35}) and (\ref{Sfi351})) do not depend upon the 
volume $V$. The function $\tilde{F}(p_{+}, p_{-}; k)$ in Eq.(\ref{Sfi351}) is simple and  
closely related to the function $\tilde{F}(p_{+}, p_{-}; k)$ defined above. The explicit form 
of this function is 
\begin{eqnarray}
 \tilde{F}(p_{+}, p_{-}; k) = \frac14 \sum_{s_{+},s_{-}} \Bigl\{ \tilde{v}(p_{+},s_{+}) 
  \Bigl[ \gamma^{\mu} \; \frac{- p_{+} + k + m_0}{2 \; k \cdot p_{+}} \; \gamma^{0} + 
  \gamma^{0} \; \frac{p_{-} - k + m_0}{2 \; k \cdot p_{-}} \; \gamma^{\mu} \Bigr] \; 
  \nonumber \\
 u(p_{-},s_{-}) \tilde{u}(p_{-},s_{-}) \Bigl[ \gamma^{0} \; \frac{- p_{+} + k + m_0}{2 \; 
 k \cdot p_{+}} \; \gamma_{\mu} + \gamma_{\mu} \; \frac{p_{-} - k + m_0}{2 \; k \cdot 
 p_{-}} \; \gamma^{0} \Bigr] v(p_{+},s_{+}) \Bigr\} \; \; , \; \label{Sfi352}
\end{eqnarray}
where we have used the facts that the photon polarization is represented in the form $\epsilon 
= \varepsilon_{\mu} \gamma^{\mu} = \varepsilon^{\mu} \gamma_{\mu}$. Here and below for any 
expression which contains a pair of identical (or repeated) indexes, where one index is 
covariant and another is contravariant, we mean summation over this `dummy' index. Also, in 
Eq.(\ref{Sfi352}) the two four-dimensional scalar products in the denominators are $k \cdot 
p_{-} = \omega \; E_{-} - {\bf k} \cdot {\bf p}_{-} = \omega (E_{-} - \mid {\bf p}_{-} \mid 
\cos\theta_{-})$ and $k \cdot p_{+} = \omega \; E_{+} - {\bf k} \cdot {\bf p}_{+} = \omega 
(E_{+} - \mid {\bf p}_{+} \mid \cos\theta_{+})$. They are traditional scalar angular-dependent 
factors (see, e.g., \cite{Heitl}, \cite{AB}). Here and everywhere below in this study, the 
notations $\theta_{+}$ and $\theta_{-}$ stand for the polar angles of ${\bf p}_{+}$ and 
${\bf p}_{-}$ vectors in respect to the direction ${\bf k}$ of the outgoing photon. The 
formulas derived above allow us to perform the final steps of QED calculations which are 
described in the next Section. 

The two sums over the spins of positron $s_{+}$ and electron $s_{-}$ in Eq.(\ref{Sfi352}) are 
determined by using the two following (Casimir) formulas \cite{Cas}: 
\begin{eqnarray}
 \sum_{s_{+}} v_{\alpha}(p_{+},s_{+}) \tilde{v}_{\beta}(p_{+},s_{+}) = - \Bigl(\frac{- 
 p_{+} + m_{0}}{2 m_{0}}\Bigr)_{\alpha\beta} = \Bigl(\frac{p_{+} - 
 m_{0}}{2 m_{0}}\Bigr)_{\alpha\beta} \; \; \label{Cas1}
\end{eqnarray}
for unpolarized positrons, and  
\begin{eqnarray}
 \sum_{s_{-}} u_{\alpha}(p_{-},s_{-}) \tilde{u}_{\beta}(p_{-},s_{-}) = \Bigl(\frac{p_{-} 
 + m_{0}}{2 m_{0}}\Bigr)_{\alpha\beta} \; \; \label{Cas2}
\end{eqnarray}
for unpolarized electrons, respectively. By using these formulas one obtains the following 
expression 
\begin{eqnarray}
 \tilde{F}(p_{+}, p_{-}; k) = \Bigl(\frac{1}{64 m^{2}_{0}}\Bigr) \; Tr\Bigl\{ (p_{+} 
 - m_{0}) \Bigl[ \gamma^{\mu} \; \frac{- p_{+} + k + m_0}{k \cdot p_{+}} \; \gamma^{0} 
 + \gamma^{0} \; \frac{p_{-} - k + m_0}{k \cdot p_{+}} \; \gamma^{\mu} \Bigr] \; 
 \nonumber \\
 (p_{-} + m_{0}) \Bigl[ \gamma^{0} \; \frac{- p_{+} + k + m_0}{k \cdot p_{+}} \; 
 \gamma_{\mu} + \gamma_{\mu} \; \frac{p_{-} - k + m_0}{k \cdot p_{-}} \; \gamma^{0} 
 \Bigr]\Bigr\} \; \; , \; \label{Sfi353}
\end{eqnarray}
where the notation $Tr$ stands for the trace of any product of four-dimensional matrices. which 
may explicitly include $\gamma-$matrices. The final expression for the unpolarized cross section 
derived in Eq.(\ref{Sfi351}) is written in the following compact form     
\begin{eqnarray}
 d\sigma = \frac{\pi Q^{2} \alpha^{3}}{4} \; \Theta(\omega - E_{-} - E_{+}) \; \frac{\omega \; 
 d\omega \; d{\bf n}_{f}}{\mid {\bf p}_{+} \mid \mid {\bf p}_{-} \mid} \; \; \frac{1}{\mid 
 {\bf q} \mid^{4}} \; F(p_{+}, p_{-}; k) \; \; , \; \label{Sfi354} 
\end{eqnarray}
where the function $F(p_{+}, p_{-}; k)$ is  
\begin{eqnarray}
  F(p_{+}, p_{-}; k) &=& Tr\Bigl\{\Bigl[ \gamma^{\mu} \; \frac{- p_{+} + k + m_0}{k \cdot 
  p_{+}} \; \gamma^{0} + \gamma^{0} \; \frac{p_{-} - k + m_0}{k \cdot p_{-}} \; \gamma^{\mu} 
  \Bigr] \; (p_{-} + m_{0}) \nonumber \\
 & &\Bigl[ \gamma^{0} \; \frac{- p_{+} + k + m_0}{k \cdot p_{+}} \; \gamma_{\mu} 
 + \gamma_{\mu} \; \frac{p_{-} - k + m_0}{k \cdot p_{-}} \; \gamma^{0} \Bigr] (p_{+} - m_{0}) 
 \Bigr\} \; \; . \; \label{Sfi354}
\end{eqnarray} 
Here we performed a cyclic permutation of matrices under the $Tr$ sign which is allowed operation in 
the matrix algebra (see, e.g., \cite{Halmos}, \$ 55).   

\section{Calculations of traces} 

In Section II we have reduced the original problem to analytical calculations of the traces of some 
four-dimensional matrices mentioned in Eq.(\ref{Sfi354}). In this Section we need to calculate these 
traces explicitly. It is clear that the formula Eq.(\ref{Sfi354}), can be represented in the form of 
three-term sum 
\begin{eqnarray}
 F = \frac{1}{(k \cdot p_{-})^{2}} \; F_1 + \frac{1}{(k \cdot p_{+})^{2}} \; F_2 + \frac{1}{(k \cdot 
 p_{+}) (k \cdot p_{-})} \; (F_3 + F_4)\; \; , \; \; \label{Sfi355}
\end{eqnarray} 
where the `partial' functions $F_{1}(p_{+}, p_{-}; k), F_{2}(p_{+}, p_{-}; k), F_{3}(p_{+}, p_{-}; k)$ 
and $F_{4}(p_{+}, p_{-}; k)$ are: 
\begin{eqnarray}
 F_{1} &=& Tr\Bigl[ \gamma^{\mu} \; (- p_{+} + k + m_0) \; \gamma^{0} \; (p_{-} + m_{0}) \; \gamma^{0} \; 
 (- p_{+} + k + m_0) \; \gamma_{\mu} \; (p_{+} - m_{0}) \Bigr] \; \; , \; \; \label{Trace1} \\
 F_{2} &=& Tr\Bigl[ \gamma^{0} \; (p_{-} - k + m_0) \; \gamma^{\mu} \; (p_{-} + m_{0}) \; \gamma_{\mu} \; 
 (p_{-} - k + m_0) \; \gamma^{0} \; (p_{+} - m_{0}) \Bigr]
\; \; \; , \; \; \label{Trace2} \\
 F_{3} &=& Tr\Bigl[ \gamma^{0} \; (p_{-} - k + m_0) \; \gamma^{\mu} \; (p_{-} + m_{0}) \; 
 \gamma^{0} \; (- p_{+} + k + m_0) \; \gamma_{\mu} \; (p_{+} - m_{0}) \Bigr] \; \; \; , \; 
 \; \label{Trace3} \\
 F_{4} &=& Tr\Bigl[ \gamma^{\mu} \; (- p_{+} + k + m_0) \; \gamma^{0} \; (p_{-} + m_{0}) \; 
 \gamma_{\mu} \; (p_{-} - k + m_0) \; \gamma^{0} \; (p_{+} - m_{0}) \Bigr] \; \; \; , \; 
 \; \label{Trace4} 
\end{eqnarray}
In these terms (or traces) we determine only those terms which contain even powers of $m_{0}$ and 
even powers of Dirac's $\gamma-$matrices. All terms with odd powers of $m_{0}$ contain odd number 
of $\gamma-$matrices. Therefore, their traces and overall contributions into the final expression 
equal zero identically. Actual calculations of the first two terms (see, Eqs.(\ref{Trace1}) and 
(\ref{Trace2})), are relatively easy. Indeed, we can write for the first term 
\begin{eqnarray}
 F_{1} &=& - 2 \; Tr\Bigl[ (- p_{+} + k + m_0) \; (\tilde{p}_{-} + m_{0}) \; (- p_{+} + k + m_0) 
 \; (p_{+} + 2 m_{0}) \Bigr] \; \; \nonumber \\ 
 &=& 2 \; Tr\Bigl[(- p_{+} + k + m_0) \; ({p}_{-} - 2 (p_{-} \cdot \gamma^{0}) \gamma^{0} - 
 m_{0}) \; (- p_{+} + k + m_0) \; (p_{+} + 2 m_{0}) \Bigr] \; \; \; , \; \; \label{Trace1A} 
\end{eqnarray}
where we have used the four following identities: $\gamma^{\mu} \; v \; \gamma_{\mu} = - 2 \; v \; , 
\; \gamma^{0} v \gamma^{0} = \tilde{v} = - v + 2 (v \cdot \gamma^{0}) \gamma^{0} \; \; , \; \; 
\gamma^{\mu} s \gamma_{\mu} = 4 s$ and $\gamma^{0} s \gamma^{0} = s \;$. These identities are obeyed 
for an arbitrary four-vector $v$ and scalar $s$. Derivation of the explicit formula for the second 
term $F_{2}$ in Eq.(\ref{Sfi355}) is similar and we do not wish to explain all details. The final 
expression is written in the form  
\begin{eqnarray}
 F_{2} = 2 \; Tr\Bigl[ (p_{-} - k + m_0) \; (p_{-} - 2 m_{0}) \; (p_{-} - k + m_0) \; (\tilde{p}_{+} 
 - m_{0}) \Bigr] \; \; , \; \; \label{Trace2A} 
\end{eqnarray}
where $\tilde{p}_{+} = - p_{+} + 2 (p_{+} \cdot \gamma^{0}) \gamma^{0}$. 

Analytical calculations of the third $F_{3}$ and fourth $F_{4}$ terms are more difficult. The central 
point here is to determine the $\gamma^{\mu} \; (p_{-} + m_{0}) \; \gamma^{0} \; (p_{-} - k + m_0) \; 
\gamma_{\mu}$ and $\gamma^{\mu} \; (- p_{+} + k + m_0) \; \gamma^{0} \; (p_{-} + m_{0}) \; 
\gamma_{\mu}$ products, which are included in the third and fourth terms, respectively. These values
are obtained by using the following matrix identities
\begin{eqnarray}
 \gamma^{\mu} \gamma_{\mu} = 4 \; \; , \; \; \gamma^{\mu} a \gamma_{\mu} = - 2 \; a \; \; , \; \;
 \gamma^{\mu} a \; b \; \gamma_{\mu} = 4 \; (a \cdot b) \; \; , \; \;  
 \gamma^{\mu} a \; b \; c \gamma_{\mu} = - 2 \; c \; b \; a  \; \; . \; \; \label{abc} 
\end{eqnarray}
From these identities one finds another general equality which is useful in our case 
\begin{equation}
 \gamma^{\mu} \; (a + \lambda m_{0}) \; \gamma^{0} \; (b + \kappa m_{0}) \gamma_{\mu} = - 2 \; b 
 \gamma^{0} \; a + 4 \; m_{0} (\kappa \; a^{0} + \lambda \; b^{0}) - 2 \; m^{2}_{0} \; \lambda \; 
 \kappa \gamma^{0} \; \; , \; \; \label{abc1} 
\end{equation} 
where $a^{0} = (a \cdot \gamma^{0}) = (\gamma^{0} \cdot a)$ and $b^{0} = (b \cdot \gamma^{0}) = 
(\gamma^{0} \cdot b)$. Here $a$ and $b$ are arbitrary four-vectors, while $\lambda$ and $\kappa$ are 
arbitrary scalars. We also applied a number of well known identities for the $\gamma-$matrices, e.g.,  
$\gamma^{\mu} \; a \; \gamma^{0} \; \gamma_{\mu} = 4 (a \cdot \gamma^{0}), \gamma^{\mu} \; \gamma^{0} 
\; \gamma_{\mu} = - 2 \; \gamma^{0}$, etc. This allows one to obtain the two explicit expressions,
which are the parts of $F_{3}$ and $F_{4}$ functions, respectively: 
\begin{eqnarray}
 & & \gamma^{\mu} \; (p_{-} + m_{0}) \; \gamma^{0} \; (- p_{+} + k + m_0) \; \gamma_{\mu} = 2 \; (p_{+} 
 - k) \; \gamma^{0} \; p_{-} + 4 \; m_{0} (- p^{0}_{+} + k^{0} - p^{0}_{-}) \nonumber \\
 &-& 2 \; m^{2}_{0} \; \gamma^{0} = 2 \; (p_{+} - k) \; \gamma^{0} \; p_{-} + 4 \; m_{0} \; (\omega - 
 E_{+} - E_{+}) - 2 \; m^{2}_{0} \; \gamma^{0} \; \label{AMF1} 
\end{eqnarray}
and 
\begin{eqnarray}
 \gamma^{\mu} \; (- p_{+} + k + m_0) \; \gamma^{0} \; (p_{-} + m_{0}) \gamma_{\mu} = 2 \; p_{-}  
 \gamma^{0} \; (p_{+} - k) + 4 m_{0} (\omega - E_{+} - E_{+}) - 2 m^{2}_{0} \gamma^{0} \; 
 \label{AMF2}  
\end{eqnarray}
By using these two identities we derive the following formulas for the $F_{3}$ and $F_{4}$ functions 
\begin{eqnarray}
 F_{3} &=& 2 \; \Bigl\{\Bigl[(p_{+} - k) \; \gamma^{0} \; p_{-} - 2 \; m_{0} \; (p^{0}_{+} - k^{0} 
 + p^{0}_{-}) + 2 \gamma^{0} \; m^{2}_{0} \Bigr] \; (p_{+} - m_{0}) \nonumber \\
 & & \gamma^{0} \; (p_{-} - k + m_{0}) \Bigr\} = 2 \; \Bigl\{\Bigl[(p_{+} - k) \; \gamma^{0} 
 \; p_{-} + 2 \; m_{0} \; (\omega - E_{+} - E_{-}) \nonumber \\
 & & + \; 2 \gamma^{0} \; m^{2}_{0} \Bigr] \; (p_{+} - m_{0}) \; \gamma^{0} \; (p_{-} - k + 
 m_{0})\Bigr\} \; \; , \; \; \label{AMF3} \\
 F_{4} &=& 2 \; Tr\Bigl\{\Bigl[(p_{+} - k) \; \gamma^{0} \; p_{-} + 2 \; m_{0} \; (\omega - E_{+} 
  - E_{+}) - \; m^{2}_{0} \; \gamma^{0} \Bigr] \; (p_{-} - k \nonumber \\
 & & + m_0) \; \gamma^{0} \; (p_{+} - m_{0}) \Bigr\} \; \; , \; \; \label{AMF4} 
\end{eqnarray}
where $p^{0}_{+} = (\gamma^{0} \cdot p_{+}) = E_{+} \; , \; p^{0}_{-} = (\gamma^{0} \cdot p_{-}) 
= E_{-} \;$ and $\; k^{0} = (\gamma^{0} \cdot k) = \omega \;$ (see above). Note that during 
one-photon annihilation at the infinitely heavy atomic nucleus the factor $\omega - E_{-} - E_{+} 
= 0$ equals zero identically. 

At the next step of our QED procedure we have to determine the actual traces of products of various 
four-dimensional matrices. This step is pretty standard and straightforward in Quantum Electrodynamics 
and for us there is no need to disclose details of these calculations. Instead, we just present the 
final analytical formulas for the $F_{1}, F_{2}, F_{3}$ and $F_{4}$ terms mentioned above. The terms 
$F_{1}$ and $F_{2}$ are  
\begin{eqnarray}
 F_{1}&=& - \; 2 \; Tr\Bigl[ \Bigr( - p_{+} p_{-} - p_{+} m_{0} + k \tilde{p}_{-} + m_{0} k - m_{0} 
 \tilde{p}_{-} - m^{2}_{0}\Bigl) \; \Bigl( - p_{+} p_{+} - 2 m_{0} p_{+}  \; \; \; \nonumber \\
 &+& k p_{+} + 2 m_{0} k + m_{0} p_{+} + 2 m^{2}_{0} \Bigl) \Bigr] \; \; \; \nonumber \\
 &=& - 16 \Bigl[ (k \cdot p_{+}) (k \cdot \tilde{p}_{-}) - m^{2}_{0} (k \cdot p_{+}) - m^{2}_{0} 
  (p_{+} \cdot \tilde{p}_{-}) - m^{2}_{0} (k \cdot \tilde{p}_{-}) - m^{4}_{0}) \Bigr] \; \; , \; 
  \label{F1}   
\end{eqnarray}
and 
\begin{eqnarray}
 F_{2}&=& - 2 \; Tr\Bigl[ \Bigr( p_{-} p_{-} - 2 p_{-} m_{0} - k {p}_{-} + 2 m_{0} k + m_{0} 
 p_{-} - m^{2}_{0}\Bigl) \; \Bigl( p_{-} \tilde{p}_{+} - p_{-} m_{0} \; \; \; \nonumber \\
 &-& k \tilde{p}_{+} + m_{0} k + m_{0} \tilde{p}_{+} - m^{2}_{0} \Bigl) \Bigr] \; \; \; 
 \nonumber \\
 &=& - 16 \Bigl[ (k \cdot p_{-}) (k \cdot \tilde{p}_{+}) + m^{2}_{0} (k \cdot \tilde{p}_{+}) - 
  m^{2}_{0} (p_{-} \cdot \tilde{p}_{+}) + m^{2}_{0} (k \cdot p_{-}) - m^{4}_{0} \Bigr] \; \; \; 
 , \; \label{F2}   
\end{eqnarray}
respectively. There is an obvious pairwise symmetry between the formulas for the $F_1$ and $F_2$ terms. 
The sum of these two terms equals 
\begin{eqnarray}
 F_1 + F_2 &=& 32 \; \Bigl\{ m^{4}_{0} + m^{2}_{0} \; [(k \cdot p_{+}) + (k \cdot p_{-}) - 
 \omega^{2}] - m^{2}_{0} \; [(p_{+} \cdot p_{-}) - 2 \; E_{+} \; E_{-}] \nonumber \\
 &+& (k \cdot p_{-}) (k \cdot p_{+}) - 2 \omega [ E_{+} (k \cdot p_{-}) + E_{-} (k \cdot p_{+})] 
 \Bigr]\Bigr\} \; \; . \; \; \label{F1F2}   
\end{eqnarray} 
Note that this expression does not contain any of the $\tilde{a}$ vectors. 

Analogous formulas for the $F_{3}$ and $F_{4}$ terms take the form 
\begin{eqnarray}
 F_3 &=& 8 \; \Bigl[(k \cdot p_{+}) (p_{-} \cdot p_{+}) + (k \cdot p_{-}) (p_{-} \cdot p_{+}) 
 - (p_{-} \cdot p_{+})^{2} + (p_{-} \cdot \tilde{p}_{-}) (p_{+} \cdot \tilde{p}_{+}) \nonumber \\
 &-& 2 \; E^{2}_{+} \; (k \cdot \tilde{p}_{-}) - 2 \; E^{2}_{-} \; (k \cdot \tilde{p}_{+}) + 2 \; 
 m^{2}_{0} \; (p_{-} \cdot p_{+}) + 2 \; m^{2}_{0} \; \Bigl((E_{-} - E_{+})^{2} - \omega^{2}\Bigr) 
 \nonumber \\
 &-& (p_{-} \cdot \tilde{p}_{+})^{2} + (p_{-} \cdot \tilde{p}_{+}) (k \cdot \tilde{p}_{+}) + 
 (p_{-} \cdot \tilde{p}_{+}) (k \cdot \tilde{p}_{-}) - m^{4}_{0} \Bigr] \; \; \; \label{F3}
\end{eqnarray}
and 
\begin{eqnarray}
 F_4 &=& 8 \; \Bigl[(k \cdot p_{+}) (p_{-} \cdot p_{+}) + (k \cdot p_{-}) (p_{+} \cdot p_{-}) 
 - (p_{-} \cdot p_{+})^{2} + (p_{-} \cdot \tilde{p}_{-}) (p_{+} \cdot \tilde{p}_{+}) \nonumber \\
 &-& 2 \; E^{2}_{+} \; (k \cdot \tilde{p}_{-}) - 2 \; E^{2}_{-} \; (k \cdot \tilde{p}_{+}) + 2 \; 
 m^{2}_{0} \; (p_{+} \cdot p_{-}) + 2 \; m^{2}_{0} \; \Bigl((E_{-} - E_{+})^{2} - \omega^{2}\Bigr) 
 \nonumber \\
 &-& (p_{-} \cdot \tilde{p}_{+})^{2} + (p_{-} \cdot \tilde{p}_{+}) (k \cdot \tilde{p}_{+}) + 
 (p_{-} \cdot \tilde{p}_{+}) (k \cdot \tilde{p}_{-}) - m^{4}_{0} \Bigr] \; \; . \; \label{F4}
\end{eqnarray}
Again, we have to note that there is an obvious pairwise symmetry between the formulas for the $F_3$ 
and $F_4$ terms. Sources of such a symmetry are discussed below. 

Now, we can determine the total sum $F$ of these four $F_1, F_2, F_3$ and $F_4$ terms and obtain the 
final answer. However, before this step, we have to discuss calculations of the factor $\mid {\bf q} 
\mid^{4}$, which is included in the denominator of the formula, Eq.(\ref{Sfi354}). Furthermore, the 
closely related ${\bf q}^{2}$ factor will also be included in the final formula. Here the 
three-dimensional vector ${\bf q} = {\bf k} - {\bf p}_{+} - {\bf p}_{-}$ is the vector of momentum 
transfer (to the nucleus). The ${\bf q}^{2}$ factor is calculated as follows 
\begin{eqnarray}
  {\bf q}^{2} &=& {\bf k}^{2} + {\bf p}^{2}_{+} + {\bf p}^{2}_{-} + 2 ({\bf p}_{+} \cdot 
  {\bf p}_{-}) - 2 ({\bf k} \cdot {\bf p}_{-}) - 2 ({\bf k} \cdot {\bf p}_{+}) \nonumber  \\
  &=& - 2 m^{2}_{0} + 2 \; (k \cdot p_{+}) + 2 (k \cdot p_{-}) - 2 (p_{-} \cdot p_{+}) 
  + \omega^{2} + E^{2}_{+} + E^{2}_{-} - 2 \omega^{2} + 2 E_{+} E_{-} \nonumber \\
 &=& - 2 m^{2}_{0} + 2 \; (k \cdot p_{+}) + 2 (k \cdot p_{-}) - 2 (p_{-} \cdot p_{+}) = 
 - q^{2} \; \; \; , \; \; \label{q2}
\end{eqnarray}  
where $p_{+}, p_{-}, k$ and $q$ are the corresponding four-vectors and $\omega = E_{+} + E_{-}$. 
In general, the four-dimensional $q^{2}$ scalar is a relativistic invariant, while similar 
three-dimensional ${\bf q}^{2}$ scalar is not. However, for some processes, e.g., for one-photon 
annihilation of the $(e^{-},e^{+})-$pair at heavy nuclei (when our additional condition $\omega = 
E_{+} + E_{-}$ is obeyed) the scalar ${\bf q}^{2}$ is a true relativistic invariant. Also from 
Eq.(\ref{q2}) one finds that $\mid {\bf q} \mid^{4} = \mid 2 \; m^{2}_{0} - 2 \; (k \cdot p_{+}) 
- 2 (k \cdot p_{-}) + 2 (p_{-} \cdot p_{+}) \mid^{2} = q^{4}$.

The product of the total sum $F$ of the four terms ($F = F_1 + F_2 + F_3 + F_4$) and 
$\frac{1}{\mid {\bf q} \mid^{4}}$ factor is written in the form 
\begin{eqnarray}
 &&8 \cdot \Pi(p_{+}, p_{-}; k) = \frac{1}{\mid {\bf q} \mid^{4} \; (k \cdot p_{+})^{2} 
 (k \cdot p_{-})^{2}} \; F(p_{+}, p_{-}; k) = \frac{8}{\mid {\bf q} \mid^{4} 
 (k \cdot p_{+})^{2} (k \cdot p_{-})^{2}} \; \nonumber \\
 &&\Bigl[\Bigl((k \cdot p_{+})^{2} + (k \cdot p_{-})^{2}\Bigr) \; \Bigl( m^{2}_{0} \; q^{2} 
 - 2 \; (k \cdot p_{+}) (k \cdot p_{-})\Bigr) - 4 \; m^{2}_{0} \; \Bigl((k \cdot p_{+}) 
 \; E_{+} + (k \cdot p_{-}) \; E_{-}\Bigr)^{2} \nonumber \\
 && - 2 \; (k \cdot p_{+}) (k \cdot p_{-}) \; q^{2} \Bigl( E^{2}_{+} +  E^{2}_{-} + (p_{+} 
 \cdot p_{-})\Bigr)\Bigr] \; \; \; . \; \; \label{FAnsw}
\end{eqnarray} 
Substitution of this expression into Eq.(\ref{Sfi354}) leads us to the final formula for the 
differential cross section of one-photon annihilation of the $(e^{-}, e^{+})-$pair of unpolarized 
electron and positron:
\begin{eqnarray}
 & &d\sigma = 2 \pi Q^{2} \alpha^{3} \; \Theta(\omega - E_{-} - E_{+}) \; \frac{\omega \; 
 d\omega \; d{\bf n}_{f}}{\mid {\bf p}_{+} \mid \mid {\bf p}_{-} \mid} \; \Pi(p_{+}, p_{-}; k) 
 \nonumber \\
 & & = 2 \pi Q^{2} \alpha^{3} \; \Theta(\omega - E_{-} - E_{+}) \; \frac{\omega \; d\omega \; 
 d{\bf n}_{f}}{\mid {\bf p}_{+} \mid \mid {\bf p}_{-} \mid} \; \frac{1}{q^{4} (k \cdot p_{+})^{2} 
 (k \cdot p_{-})^{2}} \; \Bigl[\Bigl((k \cdot p_{+})^{2} + (k \cdot p_{-})^{2}\Bigr) \; \times 
 \nonumber \\ 
 & & \; \Bigl( m^{2}_{0} \; q^{2} - 2 \; (k \cdot p_{+}) (k \cdot p_{-})\Bigr) - 4 \; m^{2}_{0} 
 \; \Bigl((k \cdot p_{+}) \; E_{+} + (k \cdot p_{-}) \; E_{-}\Bigr)^{2} - 2 \; (k \cdot p_{+}) 
 (k \cdot p_{-}) \times \nonumber \\
 & & \; q^{2} \Bigl( E^{2}_{+} +  E^{2}_{-} + (p_{+} \cdot p_{-})\Bigr)\Bigr] \; \; . \; 
 \label{Sfi355} 
\end{eqnarray}
This formula is the main result of this study. In the following Section this formula is re-written 
in a slightly different form which are more convenient to investigate some of its properties. 

To conclude this Section we note that there is an obvious similarity between one-photon annihilation 
of the $(e^{-}, e^{+})-$pair in the central Coulomb field of an infinitely heavy atomic nucleus and 
bremsstrahlung (see, e.g., \cite{AB}, \cite{Grein}). The only difference is simple and obvious, 
since  formally we have to replace the outgoing electron (for bremsstrahlung) by an incoming positron 
(for annihilation). Such a similarity can be found between our formulas (see above) and expressions, 
which were derived in \cite{BetHeit} (see, also \cite{Heitl}) for bremsstrahlung at arbitrary 
energies of the initial electron. To this point of our study we did not use such a similarity between 
our one-photon annihilation and bremsstrahlung (to avoid mistakes), but it is extensively applied in 
the next two Sections, where we consider the angular distribution of the emitted photons (or 
$\gamma-$quanta) and low-energy limit of one-photon annihilation, respectively.  

\section{Angular distribution of the emitted photons} 

The formulas produced above can be used to describe and analyze the one-photon annihilation of the 
electron-positron pair in the central Coulomb field of an infinitely heavy atomic nucleus. The 
energies of colliding light particles, i.e., electron and positron, can be arbitrary. In reality, 
accurate experiments to detect and investigate one-photon annihilation at the atomic nuclei are 
difficult to perform, since it is a three-particle process, but results of such experiments, i.e., 
formation of a high-energy photon, are neither very promising, nor illuminating. Nevertheless, angular 
correlations between two light particles $(E_{+}, {\bf p}_{+}), (E_{-}, {\bf p}_{-})$ and photon 
$(\omega, {\bf k})$ which annihilate and arise during this process are of interest for experimental 
physicists. Furthermore, one-photon annihilation of the $(e^{-}, e^{+})-$pair in the central Coulomb 
field of a heavy, positively charged $Q e$ atomic nucleus from the bound states of few-body, positron 
containing atomic systems is of great interest in many applications, including astrophysics and solid 
state physics. Similar annihilation of the $(e^{-}, e^{+})-$pair corresponds to the low-energy limit. 
In this and next Sections we briefly discuss these two problems, i.e., angular correlations and 
low-energy limit for the formula, Eq.(\ref{Sfi355}), derived above.    

First, we note that our final formula for the differential cross section of one-photon annihilation, 
Eq.(\ref{Sfi355}), is written in a manifestly covariant form in the four-dimensional space-time, or 
in other words, this expression is a true relativistic invariant. However, in experimental physics 
it is better to use the traditional three-dimensional geometry. Therefore, we have to deal with the 
trigonometric functions of various angles. For our one-photon annihilation the angular problem is 
formulated as follows. There are two original particles (electron $e^{-}$ and positron $e^{+}$) and 
one final photon $k$ which take part in our process. This means that we have three independent 
$3D-$vectors associated with these particles. These vectors are ${\bf p}_{+}, {\bf p}_{-}$ and 
${\bf k}$, respectively. In addition to these vectors we have an infinitely heavy, central nucleus 
$(+ Q e)$ which also plays a substantial role in our process. In our affine three-dimensional 
vector-space we can always assume that the vectors ${\bf p}_{+}, {\bf p}_{-}$ and ${\bf k}$ begin 
from the heavy atomic nucleus, or in other words, they originate at the nucleus. Our goal below is 
to derive the explicit formula which describes angular correlations between these three vectors for 
different energies. 

The four-dimensional scalar products $(k \cdot p_{+})$ and $(k \cdot p_{-})$ are written in the 
forms: $(k \cdot p_{+}) = \omega (E_{+} - \mid {\bf p}_{+} \mid \cos\theta_{+})$ and $(k \cdot 
p_{-}) = \omega (E_{-} - \mid {\bf p}_{-} \mid \cos\theta_{-})$. To obtain similar formula for 
the four-dimensional scalar product $(p_{+} \cdot p_{-})$ one has to apply the formula known from 
tetrahedron geometry (see, e.g., \cite{Mod}): 
\begin{eqnarray}
 (p_{+} \cdot p_{-}) &=& E_{+} E_{-} - \mid {\bf p}_{+} \mid \mid {\bf p}_{-} \mid \cos\theta_{+-} 
 = E_{+} E_{-} - \mid {\bf p}_{+} \mid \mid {\bf p}_{-} \mid (\cos\theta_{+} \cos\theta_{-} 
 \nonumber \\
  &+& \sin\theta_{+} \sin\theta_{-} \cos\phi) \; \; , \; \label{tetra1} 
\end{eqnarray}
where $\theta_{+}$ and $\theta_{-}$ are the polar angles of ${\bf p}_{+}$ and ${\bf p}_{-}$ vectors 
in respect to the direction ${\bf k}$ of the emitted photon (see above), while $\theta_{+-}$ is the 
angle between the positron and electron three-dimensional momenta, i.e., between ${\bf p}_{+}$ and 
${\bf p}_{-}$ vectors. The angle $\phi$ equals to the angle between the two planes which are formed 
by the two pairs of vectors: $({\bf k}, {\bf p}_{+})$ and $({\bf k}, {\bf p}_{-})$, respectively. 
From here one easily finds the following expression  
\begin{eqnarray}
 q^{2} = - {\bf q}^{2} &=& - 2 m^{2}_{0} + 2 \; \omega \; (E_{+} - \mid {\bf p}_{+} \mid \; 
 \cos\theta_{+}) + 2 \; \omega \; (E_{-} - \mid {\bf p}_{-} \mid \; \cos\theta_{-}) \nonumber \\
 &-& 2 \Bigl[ E_{+} E_{-} - \mid {\bf p}_{-} \mid \mid {\bf p}_{+} \mid (\cos\theta_{+} \cos\theta_{-} 
 \; + \; \sin\theta_{+} \sin\theta_{-} \cos\phi) \Bigr] \; \; . \; \label{q2cos} 
\end{eqnarray}
and $\mid {\bf q}^{4} \mid = \mid (- {\bf q}^{2}) (- {\bf q}^{2}) \mid$. 

After a few relatively simple, additional transformations in Eq.(\ref{Sfi355}) one obtains the following 
formula which describes the angular distribution of the final $\gamma-$quanta which are emitted during 
our one-photon annihilation of the electron-positron pair at the heavy atomic nucleus: 
\begin{eqnarray}
 & &d\sigma = 2 \pi Q^{2} \alpha^{3} \; \Theta(\omega - E_{-} - E_{+}) \; \frac{d\omega \; d{\bf 
 n}_{f}}{\omega \mid {\bf p}_{+} \mid \mid {\bf p}_{-} \mid} \; \frac{1}{\mid {\bf q} 
 \mid^{4}} \; \Bigl[ (4 \; E^{2}_{-} - {\bf q}^{2}) \; \times \nonumber \\
 & & \frac{{\bf p}^{2}_{+} \; \sin^{2}\theta_{+}}{(E_{+} - \mid {\bf p}_{+} \mid \cos\theta_{+})^{2}} 
 + (4 \; E^{2}_{+} - {\bf q}^{2}) \frac{{\bf p}^{2}_{-} \; \sin^{2}\theta_{-}}{(E_{-} - \mid 
 {\bf p}_{-} \mid \cos\theta_{-})^{2}} \; \nonumber \\
 & & + 2 \; \omega^{2} \; \frac{{\bf p}^{2}_{+} \; \sin^{2}\theta_{+} + {\bf p}^{2}_{-} \; 
 \sin^{2}\theta_{-}}{(E_{+} - \mid {\bf p}_{+} \mid \cos\theta_{+}) (E_{-} - \mid 
 {\bf p}_{-} \mid \cos\theta_{-})} \; \nonumber \\
& & - 2 \; ( 2 E^{2}_{+} + 2 E^2_{-} - {\bf q}^{2} ) \; 
\frac{({\bf p}_{+} \cdot {\bf p}_{-}) - \mid {\bf p}_{+} \mid \mid {\bf p}_{-} \mid \; \cos\theta_{+} 
\; \cos\theta_{-}}{(E_{+} - \mid {\bf p}_{+} \mid \cos\theta_{+}) (E_{-} - \mid {\bf p}_{-} \mid 
\cos\theta_{-})} \Bigr] \; \; . \; \label{Sfi3555}
\end{eqnarray}
In the case of one-photon annihilation the formula, Eq.(\ref{Sfi3555}), plays the same role, which the 
famous Bethe-Heitler formula \cite{BetHeit} plays for description of angular distribution of 
bremsstrahlung. In fact, the formula, Eq.(\ref{Sfi3555}), looks similar to the Bethe-Heitler formula, 
but there is an important difference between these two formulas. Indeed, in the case of one-photon 
annihilation we have only two angular variables, while for bremsstrahlung one finds three angular 
variables. The formula, Eq.(\ref{Sfi3555}), is convenient for applications in actual experiments. From 
theoretical point of view the following integration in Eq.(\ref{Sfi3555}) over the angular variables 
(or angles, for short) require almost the same operations which are used to deal with the 
non-relativistic four-body systems \cite{Fro2006}. Here we cannot delve deeply into this interesting 
problem. Instead, let us obtain the low-energy limit for the cross section, Eq.(\ref{Sfi3555}). 

\section{Low-energy limit and one-photon annihilation rate}

The low-energy limit of the cross section of one-photon annihilation can be derived by using the 
following standard replacements (see, e.g., \cite{BLP}) in Eq.(\ref{Sfi3555}): $E_{\pm} \Rightarrow 
m_{0}, \omega \Rightarrow 2 m_{0}, q \Rightarrow 2 m_{0}$, etc. In this approximation one also finds 
$\omega \gg \mid {\bf p}_{\pm} \mid, E_{+} \gg \mid {\bf p}_{+} \mid$ and $E_{-} \gg \mid {\bf p}_{-} 
\mid$. After such a replacement we find that only one (third) term survives in Eq.(\ref{Sfi3555}) and 
this formula takes the new form 
\begin{eqnarray}
 d\sigma_{nr} = \frac{8 \; \pi \; Q^{2} \; \alpha^{3}}{m^{5}_{0}} \; \Theta(\omega - E_{-} - E_{+}) 
 \; \frac{{\bf p}^{2}_{+} \; \sin^{2}\theta_{+} + {\bf p}^{2}_{-} \; \sin^{2}\theta_{-}}{\mid {\bf 
 p}_{+} \mid \mid {\bf p}_{-} \mid} \; d{\bf n}_{f} \; . \; \label{Sfi356}
\end{eqnarray}
where the notation $nr$ means `non-relativistic'. The $\Theta$ function in this formula can be eliminated 
if one specify the `convenient' frame of references. For instance, we can chose the frame where $p_{-} = 
(m_{0}, {\bf p}_{-})$, while $p_{+} = (m_{0}, {\bf p}_{+}) = (m_{0}, - {\bf p}_{-})$. In this frame the 
standard calculations (see, e.g., \cite{Grein}) reduce Eq.(\ref{Sfi356}) to the form: 
\begin{eqnarray}
 d\sigma_{nr} = \frac{16 \; \pi \; Q^{2} \; \alpha^{3}}{m^{5}_{0}} \; \sin^{2} \theta_{-} \; 
 d{\bf n}_{f} = \frac{16 \; \pi \; Q^{2} \; \alpha^{3}}{m^{5}_{0}} \; \sin^{2} \theta_{-} \;  
 \cos \theta_{-} d\theta_{-} d \phi_{-} \; . \; \label{Sfi357}
\end{eqnarray} 
Finally, by performing integration over angular variables in this formula we derive the following formula 
for the non-relativistic cross section of the one-photon annihilation of one electron-positron pair at 
the central atomic nucleus: 
\begin{eqnarray}
 \sigma_{nr} = \frac{32 \pi^{2} Q^{2} \; \alpha^{3}}{3 \; m^{5}_{0}} \; = 
 \frac{32 \pi^{2} Q^{2} \; \alpha^{8}}{3} \; \; a^{2}_{0} \; \label{Sfi358}
\end{eqnarray}
where $a_{0} = \frac{\hbar^{2}}{m_{e} e^{2}} \approx 5.29177210903 \cdot 10^{-9}$ \; $cm \;$ \cite{NIST} 
is the Bohr radius. Note that the right side of the last equation is written in atomic units, where 
$\hbar = 1, m_{e} = 1$ and $e = 1$.

As mentioned above the goal of this study is to derive the explicit formula for the one-photon annihilation 
of the bound electron-positron pair in the field of atomic nuclei. For the bound $(e^{-}, e^{-})-$pair we 
need to multiply the final formula, Eq.(\ref{Sfi358}), by the electron-positron density at the atomic 
nucleus, or in other words, by the triple delta-function $\langle \delta_{+ Q -} \rangle$. The arising 
general formula takes the form: 
\begin{eqnarray}
 A = \frac{32 \; \pi^{2} \; Q^{2} \; \alpha^{8}}{3} \; m \; n \; \; \langle \delta_{+ Q -} \rangle \; \; 
 a_{0}^{-1} \; , \; \label{Sfi3580}
\end{eqnarray}
where it is assumed that this atomic system contains $m$ bound positrons and $n$ bound electrons. Note that 
the value $A$ is expressed in $cm^{-1}$, since the expectation value of triple delta-function $\langle 
\delta_{+ Q -} \rangle$ is expressed in the $a_{0}^{-3}$ units (see, e.g., \cite{Heitl}). Finally, the value 
$A$ is the inverse annihilation length. In reality, for few-body systems with bound electrons and positrons 
there is a more convenient value which is called the annihilation rate (see, e.g., \cite{AB} and earlier 
references therein). In general, there are zero-, one-, two-, three-, four- and multi-photon annihilation 
rates. Formally, these values are defined as the product of $A$ values and the velocity of light in vacuum 
$c$ = 299792458 $cm \cdot sec^{-1}$, which is written in the form $\Bigl(\frac{c}{a_{0}}\Bigr) \; a_{0}$. 
For example, for the rate of one-photon annihilation of the electron-positron pair in the Coulomb field of 
a central, infinitely heavy atomic nucleus, which is designated below as $\Gamma^{(b)}_{1 \gamma}$, one 
obtains: 
\begin{eqnarray} 
 \Gamma^{(b)}_{1 \gamma} = \frac{32 \; \pi^{2} \; Q^{2} \; \alpha^{8}}{3} \; m \; n \; 
 \langle \delta_{+ Q -} \rangle \; a_{0} \; \Bigl(\frac{c}{a_{0}}\Bigr) \; sec^{-1} \; , \; 
 \label{Sfi3585}
\end{eqnarray}
where the product of the cross section, expectation value of triple delta-function and factor $a_{0}$ is
dimensionless, while the last factor, i.e., $\Bigl(\frac{c}{a_{0}}\Bigr)$ provides the required and correct 
dimension $\; sec^{-1}$ for the annihilation rate. From now on, all expectation values in this formula 
(and similar formulas below) must be taken in atomic units, where $\hbar = 1, m_{e} = 1$ and $e = 1$. The 
formula, Eq.(\ref{Sfi3585}), is the final result of our study and it allows one to evaluate the rate of 
one-photon annihilation of the electron-positron pair in the Coulomb field of a central, infinitely heavy 
atomic nucleus. Note also that the ratio $\frac{c}{a_{0}} = 5.665256398 \cdot 10^{18}$ $sec^{-1}$ (or 
$Hertz$ ($Hz$)) is the specific (atomic) annihilation frequency, which arises in all problems related to 
annihilation of the $(e^{-}, e^{+})-$pair(s) from the bound states of atoms, ions, molecules, etc. 

\section{Applications to actual atomic systems}

Let us apply our formulas derived above to actual atomic few-body systems with bound positrons. In order 
to obtain all expectation values included in Eq.(\ref{Sfi3585}) we have to solve the non-relativistic 
Shcr\"{o}dinger equation $\hat{H} \Psi = E \Psi$ for a given few-body system. In this equation $\Psi$ is 
the bound state wave function, $\hat{H}$ is the Hamiltonian and $E$ is some negative numerical parameter 
which is called the total energy of this bound state. Without loss of generality, we can assume that the 
wave function $\Psi$ of any bound state has the unit norm, i.e., $\mid \langle \Psi \mid \Psi \rangle 
\mid^{2} = 1$. By using this wave function $\Psi$ we can determine the expectation values of all bound 
state properties of the given few-body system in any selected bound state. This includes the expectation 
values of three-particle delta-functions which are designated in our formulas above as $\langle 
\delta_{+ Q -} \rangle = \langle \Psi \mid \delta({\bf r}_{- N}) \delta({\bf r}_{+ N}) \mid \Psi \rangle$ 
and $\langle \delta_{- + -} \rangle = \langle \Psi \mid \delta({\bf r}_{+ -}) \delta({\bf r}_{- -}) \mid 
\Psi \rangle$. 

By using our formulas derived in this study we can investigate the one-photon annihilations of the 
$(e^{+}, e^{-})-$pair in the positronium hydrides HPs (or ${}^{1}$HPs), DPs (or ${}^{2}$HPs), TPs 
(or ${}^{3}$HPs). Stability of these Coulomb as the bound four-body atomic systems was experimentally 
shown in \cite{CH4} where the molecules of positronium HPs hydride were created during collisions 
between a beam of positrons and molecular methane (gas). Astrophysical interest to these systems and 
their bound state properties are discussed, e.g., in \cite{Fro2022} (see also references therein). 
In this study we need to evaluate the one-photon annihilation rate and we simply cannot discuss all 
earlier papers on positronium hydrides. By applying our formula, Eq.(\ref{Sfi3585}), where $Q = 1, 
m = 1 \; n = 2$, to the four-body positronium hydrides, we have found that $\Gamma^{(b)}_{1 
\gamma}(^{\infty}$HPs) $\approx$ 0.0828319(15) $sec^{-1}$. This result has been obtained for the 
model ${}^{\infty}$HPs system with the infinitely heavy hydrogen nucleus. Results, i.e., the 
one-photon annihilation rates $\Gamma^{(b)}_{1 \gamma}$ for analogous four-body ${}^{3}$HPs, 
${}^{2}$HPs and ${}^{1}$HPs systems are: \; 0.0827698(15) $sec^{-1}$ , \; 0.0827389(15) $sec^{-1}$ 
, \; 0.0826461(15) $sec^{-1}$, respectively. Here the notations ${}^{3}$H, ${}^{2}$H and ${}^{1}$H 
mean the nuclei of hydrogen isotopes, i.e., tritium, deuterium and protium, respectively. The 
nuclear masses of these hydrogen nuclei have been taken from \cite{Fro2022}. 

Note that in any atomic system with positrons where $m \ge 1$ and $n \ge 2$, or $m \ge 2$ and $n 
\ge 1$, another one-photon annihilation is also possible. Such an annihilation of the $(e^{-}, 
e^{+})-$pair is related with the localization of two electrons and one positron (or two positrons 
and one electron) at the same spatial point. The formula for the corresponding annihilation rate 
$\Gamma^{(a)}_{1 \gamma}$ for the three-body $e^{-} - e^{+} - e^{-}$ ion (or Ps$^{-}$ ion) has 
been derived in \cite{Kru}. For the positronium hydrides in atomic units this formula is written 
in a slightly different form \cite{OurHPs} 
\begin{eqnarray} 
 \Gamma^{(a)}_{1 \gamma} = \frac{64 \; N_{e^{-} e^{+} e^{-}} \; \pi^{2} \; \alpha^{8}}{27} 
 \; \Bigl(\frac{c}{a_{0}}\Bigr) \; \langle \delta_{- + -} \rangle \; \; sec^{-1} \; , \; 
 \label{Sfi358C}
\end{eqnarray}
where $\langle \delta_{- + -} \rangle$ is the expectation value of the triple delta-function (i.e., 
the electron-positron-electron delta-function) in atomic units, while $N_{e^{-} e^{+} e^{-}}$ is the 
total number of electron-positron triples that can be formed from all bound electrons and positrons 
in a given atomic system. Each of the three-body Ps$^{-}$ and/or Ps$^{+}$ contains one 
electron-positron triplet. Atomic system with $m \ge 2$ bound positrons and $n \ge 2$ bound 
electrons contains $N_{e^{-} e^{+} e^{-}} = m \; C^{2}_{n} + n \; C^{2}_{m} = m \; n \; \Bigr[ 
\frac{m + n }{2} - 1 \Bigl]$ such triples. Here $C^{q}_{p}$ is the number of combinations from $p$ 
by $q$ (here $p$ and $q$ are integer). The formula, Eq.(\ref{Sfi358C}), is based on the results of 
earlier papers \cite{Kru}, \cite{Nels}. Result obtained in \cite{ChuPon} is wrong by factor of 2, 
since the authors took into account only two Feynman diagrams (not four as it must be for the 
Ps${}^{-}$ ion). More details of these calculations can be found in \cite{Fro1995}. Numerical value 
of this one-photon annihilation rate equals $\approx \Gamma^{(a)}_{1 \gamma}(^{\infty}$HPs) $\approx$ 
0.0358242(15) $sec^{-1}$. Analogous one-photon rates for the four-body ${}^{3}$HPs, ${}^{2}$HPs and 
${}^{1}$HPs systems are: \; 0.0358045(15) $sec^{-1}$, \; 0.0357947(15) $sec^{-1}$, \; 0.0357651(15) 
$sec^{-1}$.

Now, let us evaluate the one-photon annihilation rate in the excited electron-triplet $2^{3}S-$state
of the helium-positron $e^{+}[^{4}$He($2^{3}S_e$)] and $e^{+}[^{4}$He($2^{3}S_e$)] ions. These ions 
are the Coulomb four-body systems each of which contains one weakly-bound positron and the triplet 
two-electron pair. Here we also consider the model positron-helium $e^{+}[^{\infty}$He($2^{3}S_e$)] 
ion with the infinitely heavy nucleus. It is clear that in calculations of these electron-triplet 
states in such systems we have to assume that $Q = 2, m = 1 \; n = 2$ in Eq.(\ref{Sfi3585}). By using 
our formula, Eq.(\ref{Sfi3585}), for the model positron-helium $e^{+}[^{\infty}$He($2^{3}S_e$)] ion 
we have found that $\Gamma^{(b)}_{1 \gamma} \approx \; 8.32043(3) \cdot 10^{-4}$ $sec^{-1}$. For the 
actual positron-helium $e^{+}[^{4}$He($2^{3}S_e$)] and $e^{+}[^{3}$He($2^{3}S_e$)] ions these 
one-photon annihilation rates are: $\; 8.30342(3) \cdot 10^{-4}$ $sec^{-1}$ and $\; 8.29841(3) \cdot 
10^{-4}$ $sec^{-1}$, respectively. The first one-photon annihilation rate $\Gamma^{(a)}_{1 \gamma}$ 
for each of these weakly-bound electron-triplet states is a very small number (almost equals zero). 
It is possible to say that, similar one-photon annihilation in this state is strictly prohibited in 
such a state. Another interesting fact follows from the direct numerical comparison of one-photon 
annihilation rates $\Gamma^{(b)}_{1 \gamma}$ in the $e^{+}[^{M}$He($2^{3}S_e$)] ions with these same 
annihilation rates the four-body HPs systems. In the case of helium-positron ions one obtains 
significantly smaller one-photon annihilation rates $\Gamma^{(b)}_{1 \gamma}$ than in the HPs systems. 
This can be explained based on very fast decrease of the expectation value of triple delta-function
(i.e., the $\langle \delta_{+ Q -} \rangle$ value) when the nuclear charge $Q e$ increases. Briefly, 
one can say that it is much more difficult for a slow, non-relativistic positron to get closer to 
that atomic nucleus which has a larger electric charge $Q e$. 

\subsection{General formula for one-photon annihilation rate of the electron-positron pair} 

By comparing the formulas, Eq.(\ref{Sfi3585}) and Eq.(\ref{Sfi358C}), for the one-photon annihilation 
rates, it is easy to note their substantial similarity. This can be used to derive the united (or 
universal) expression for one-photon annihilation rate of the $(e^{-}, e^{+})-$pair. In reality, such 
an `united' formula can be derived and its derivation is essentially based on the results of this 
study and two earlier papers \cite{Kru} and \cite{ChuPon} on one-photon annihilation of the 
electron-positron pair in the three-body Ps$^{-}$ (or Ps$^{+}$) ion. Here we cannot describe all 
details that led us to the final formula for one-photon annihilation rate (in the non-relativistic 
limit), but we will simply present and discuss it: 
\begin{eqnarray} 
 \Gamma_{1 \gamma} =  [f_0 \; f_1 \; f^{2}_2] \; \; \frac{32 \; \pi^{2} \; Q^{2} \; \alpha^{8}}{3} 
 \; \Bigl(\frac{c}{a_{0}}\Bigr) \; \langle \delta_{+ Q -} \rangle \; \; sec^{-1} \; , \; 
 \label{Sfi358F}
\end{eqnarray}
where all notations have the same sense as in Eq.(\ref{Sfi3585}) above, but this formula contains a 
product of three additional factors $f_0, f_1(m, n)$ and $f_2 = f_2(M_Q, m_e)$, where the last factor 
is 
\begin{eqnarray} 
  f_2 = f_2(M_Q, m_e) = \frac{M_Q}{M_Q + m_e + m_p} = \frac{M_Q}{M_Q + m_e + m_e} = \frac{M_Q}{M_Q 
  + 2 m_e } = \frac{M_Q}{M_Q + 2} \; , \nonumber 
\end{eqnarray}
where $M_Q$ is the mass of a `central' particle $Q$ expressed in $m_e$, while $m_p = m_e$ is the 
positron mass. The choice of the $f_0$ and $f_1$ factors in Eq,(\ref{Sfi358}) depends upon 
annihilation nature of the `central' particle $Q$. If this `central' particle is either electron, 
or positron then $f_0 = 2$ and $f_1(m, n) = N_{e^{-} e^{+} e^{-}}$, where $N_{e^{-} e^{+} e^{-}} 
= N_{e^{+} e^{-} e^{+}}$ is the total number of electron-positron triples. However, if this 
`central' particle is a some `heavy' particle (neither positron, nor electron), then $f_0 = 1$ 
and $f_1(m, n) = m \; n$. By following these simple rules we obtain for the ${}^{\infty}$HPs 
system $[f_0 \; f_1 \; f^{2}_2] = 1 \cdot m \; n \cdot 1 = m \; n = 2$ and we are back to our 
Eq.(\ref{Sfi3585}). Analogously, for the Ps$^{-}$ ion one finds $[f_0 \; f_1 \; f^{2}_2] = 2 \; 
(N_{e^{-} e^{+} e^{-}}) \; \Bigl(\frac13 \Bigr)^{2}= \frac29$ and this leads directly to 
Eq.(\ref{Sfi358C}). In these two examples we assumed that $Q = 1$. However, this our formula,  
Eq.(\ref{Sfi358F}), we have successfully tested for arbitrary bound atomic systems with positrons. 
Results of such applications will be published elsewhere. Note that in this formula, 
Eq.(\ref{Sfi358F}), we do not to assume anymore that all heavy particles are `infinitely heavy'.  

\section{Conclusion}

We have investigated the process of one-photon annihilation of the bound electron-positron pair 
in the Coulomb field of a heavy atomic nucleus. Finally, this important QED problem has been 
solved completely and accurately. It is interesting to note that one-photon annihilation of the 
bound electron-positron pair in the Coulomb field of an infinitely heavy atomic nucleus is one 
of a very few fundamental QED-tasks that are represented by the two-vertex Feynman diagrams. 
Many similar processes were extensively studied in 1930's. 

Our formulas for the differential and total cross section of one-photon annihilation of the 
bound $(e^{-}, e^{+})-$pair in the Coulomb field of infinitely heavy atomic nucleus have been 
derived in the closed analytical forms. These formulas allow one to investigate and describe 
(completely and accurately) the phenomenon of one-photon annihilation of the $(e^{-}, 
e^{+})-$pair in the Coulomb field of heavy atomic nuclei. In particular, now we can predict 
angular correlations between the original directions of incoming electron and/or positron and 
the final direction of outgoing photon (or emitted $\gamma-$quantum). We also determined the 
low-energy limit for the cross section of one-photon annihilation in the Coulomb field of a 
central atomic nucleus. By using this result we derived the explicit expression for the 
one-photon annihilation rate $\Gamma^{(b)}_{1 \gamma}$ in atomic systems which contain at 
least two electrons and one positron. This allows us to achieve the main goal of this study. 
Then based on this our formula and results from \cite{Kru} and \cite{ChuPon} we derived the 
universal formula for the one-photon annihilation rate in aa arbitrary few-body systems which 
contains one heavy particle, e.g., atomic nucleus and/or positively charged muon, two (or more) 
electrons and at least one positron. This is another advantage of our study. 

In fact, all formulas derived in this study have been applied to evaluate the rates of one-photon 
annihilation in some atomic systems each of which contains one additional positron. Such systems 
include the four-body positronium hydrides HPs and quasi-stable electron-triplet $S(L = 0)-$states 
in the positron-helium $e^{+}[^{3}$He($2^{3}S_e$)] and $e^{+}[^{4}$He($2^{3}S_e$)] ions. For none 
of these ions the rate of one-photon annihilation at the atomic nucleus has ever been determined 
in earlier studies. Finally, we have to note that the both differential and total cross section of 
one-photon annihilation of the electron-positron pair in the Coulomb field of hydrogen atomic 
nucleus, i.e., proton, exactly coincide with the corresponding cross sections of one-photon 
annihilation in the field of antihydrogen nucleus, i.e., antiproton. The same conclusion is true 
for the annihilation rates. These facts can be useful in future research of antiproton atomic 
systems with electrons and positrons. 

To conclude this paper we want to note that the current experimental limit for detection of 
$(e^{-}, e^{+})-$annihilation can be evaluated as $\approx 1 \cdot 10^{3}$ $sec^{-1}$. This 
follows from recent experiments (see, e.g., \cite{Dary} - \cite{Nagash2}, references therein 
and some other papers). This is significantly exceed the experimental limit which is needed 
for reliable detection and measuring of the one-photon annihilation rates. However, the 
actual progress in this experimental area of research is amazingly fast and, probably, 
one-photon annihilation rates in light atoms and ions  can be measured in 25 - 30 years from 
now. \\ 

\noindent 
{\bf ACKNOWLEDGMENTS} \\
This manuscript is dedicated to the memory of Professor Joachim Reinhardt (`QED person') who 
was always interested in similar QED problems. Prof. G.W.F. Drake (University of Windsor, 
Windsor, Ontario) has provided sufficient computational resources to finish this study. \\

\noindent 
{\bf Author Contributions} \\ 
Alexei M. Frolov: Conceptualization (equal); Formal analysis (equal);
Investigation (equal); Writing - original draft (equal); Writing -   
review \& editing (equal). \\

\noindent
{\bf DATA AVAILABILITY} \\
The data that support the findings of this study are generated inside of the paper, or can be 
found in the references mentioned in bibliography. \\

\noindent
{\bf AUTHOR DECLARATIONS} \\
\noindent 
{\bf Conflict of Interest} \\
The authors have no conflicts to disclose. No potential conflict of 
interest was reported by the author(s).  \\


\begin{thebibliography}{99}

\bibitem{Heitl} W. Heitler, \textit{The Quantum Theory of Radiation} (3rd ed., Oxford University 
Press, London, UK (1954)). 

\bibitem{OurHPs} A.M. Frolov and V.H. Smith, Jr., Phys. Rev. A {\bf 55}, 2662 (1997). 

\bibitem{Dra1} R.J. Drachman, Can. J. Phys. 60, 494 (1982).

\bibitem{Svens} R. Svensson, Astrophys. J. {\bf 258}, 321 (1982).

\bibitem{Dermer} C.D. Dermer and J.G. Skibo, Astrophys. J. Lett. {\bf 487}, L57 (1997).

\bibitem{Pran} N. Prantzos, Astron. Astrophys. {\bf 449}, 869 (2006).

\bibitem{Cheng} K.S. Cheng, D.O. Chernyshov, and V.A. Dogiel, Astrophys. J. 
{\bf 645}, 1138 (2006).

\bibitem{Su} M. Su, T. R. Slatyer, and D. P. Finkbeiner, Astrophys. J.{\bf 724}, 
1044 (2010). 

\bibitem{Crock} R.M. Crocker and F. Aharonian, Phys. Rev. Lett. {\bf 106}, 101102 (2011).

\bibitem{Fro2023} A.M. Frolov, Physics of Plasmas {\bf 30}, 102701 (2023).

\bibitem{Panth1} F.H. Panther, R.M. Crocker, Y. Birnboim, I.R. Seitenzahl and 
A.J. Ruiter, Mon. Not. R. Astron. Soc. {\bf 474}, L17 (2018).

\bibitem{Panth2} F.H. Panther, Galaxies {\bf 6}, 39 (2018).

\bibitem{Fro2022} A.M. Frolov, J. Mol. Phys. {\bf 120}(6), e2019337 (2022). 

\bibitem{Fro2025} A.M. Frolov, European Physical Journal D {\bf 79}, article \# 38 (2025). 

\bibitem{AB} A. Akhiezer and V.B. Berestetskii, \textit{Quantum Electrodynamics} 
(4th ed., Science, Moscow (1981)) [in Russian]. 

\bibitem{Grein} W. Greiner and J. Reinhardt, \textit{Quantum Electrodynamics} 
(4th ed., Springer Verlag, Berlin (2009)).

\bibitem{Cas} H.B.G. Casimir, Helv. Phys. Acta {\bf 6}, 287 (1933). 

\bibitem{Halmos} P.R. Halmos, \textit{Finite-Dimensional Vector Spaces} (2nd. ed., 
D. Van Nostrand Company, Inc., New York (1958)). 

\bibitem{BetHeit} H.A. Bethe and W. Heitler, Proc. Roy. Soc. A {\bf 146}, 83 (1934). 
Press, London, UK (1934)). 

\bibitem{Mod} P.S. Modenov, \textit{Analytical Geometry} (Moscow State University 
Publ., Moscow (1959)) [in Russian]. 

\bibitem{Fro2006} A.M. Frolov, J. Phys. A {\bf 39}, 15421 (2006). 

\bibitem{BLP} V.B. Berestetskii, E.M. Lifshitz and L.P. Pitaevskii, \textit{Relativistic 
Quantum Theory} (2nd ed., reprinted by Elsevier Ltd., Burlington, MA (2012)). 

\bibitem{NIST} see, e.g., https://physics.nist.gov/cgi-bin/cuu/Value? 

\bibitem{Kru} S.I. Kryuchkov, J. Phys. B: At. Mol. Phys. {\bf 27}, L61 (1994). 

\bibitem{Nels} Bent Nielsen, unpublished (1987). In 1997 I received a letter from
Richard J. Drachman which included a collection of $\approx$ 25 pages of hand made 
QED calculations for the one-photon annihilation rate in the three-body Ps$^{-}$ 
ion. The result of these calculations exactly coincided with the result from 
\cite{Kru}. On the first page it was a signature `Nielsen'. Probably, it was Bent 
Nielsen from Stony Brook University. 

\bibitem{ChuPon} M.-C. Chu and V. P\"{o}nish, Phys. Rev. C \textbf{33}, 2222 (1986). 

\bibitem{Fro1995} A.M. Frolov, S.I. Kryuchkov and V.H. Smith, Jr., Phys. Rev. A 
{\bf 51}, 4514 (1995). 

\bibitem{CH4} D.M. Schrader, F.M. Jacobsen, N.-P. Fradsen and U. Mikkelsen, Phys. Rev. 
Lett. {\bf 69}, 57 (1992). 

\bibitem{Dary} M. Emami-Razavi and J.W. Darywich, Eur. Phys. J. D {\bf 75}, 188 (2021). 

\bibitem{Bress} D. Bressanini, Phys. Rev. A {\bf 104}, 022819 (2021). 

\bibitem{Nagash2} Ya. Nagashima, K. Michishio, L. Chiari and Yu. Nagata, J. Phys. B: 
At. Mol. Phys. {\bf 54} 212001 (2021). 

\end{thebibliography}
\end{document}